\documentclass[aps,pra,twocolumn,amsmath,amssymb,nofootinbib,superscriptaddress]{revtex4}

\newcommand{\ket}[1]{|#1\rangle}

\usepackage[pdftex]{graphicx}
\usepackage{mathrsfs}
\usepackage[colorlinks]{hyperref}

\begin{document}

\bibliographystyle{apsrev}

\title{A simple scheme for universal linear optics quantum computing with constant experimental complexity using fiber-loops}

\author{Peter P. Rohde}
\email[]{dr.rohde@gmail.com}
\homepage{http://www.peterrohde.org}
\affiliation{Centre for Engineered Quantum Systems, Department of Physics and Astronomy, Macquarie University, Sydney NSW 2113, Australia}

\date{\today}

\frenchspacing

\begin{abstract}
Recently, Motes, Gilchrist, Dowling \& Rohde [Phys. Rev. Lett. \textbf{113}, 120501 (2014)] presented a scheme for photonic boson-sampling using a fiber-loop architecture. Here we show that the same architecture can be modified to implement full, universal linear optics quantum computing, in various incarnations. The scheme employs two embedded fiber-loops, a single push-button photon source, three dynamically controlled beamsplitters, and a single time-resolved photo-detector. The architecture has only a single point of interference, and thus may be significantly easier to align than other schemes. The experimental complexity of the scheme is constant, irrespective of the size of the computation, limited only by fiber lengths and their respective loss rates.
\end{abstract}

\maketitle

%
%

\section{Introduction}

Quantum computation promises to efficiently implement algorithms intractable on classical computers \cite{bib:NielsenChuang00}. Many physical realizations have been proposed, using different physical systems to represent qubits and their evolution. Linear optics quantum computing (LOQC) \cite{bib:KokLovett}, originally proposed by Knill, Laflamme \& Milburn (KLM) \cite{bib:KLM01}, is one of the most promising proposals, owing to the inherently long decoherence times of photons, and the relative ease with which to prepare, evolve, and measure them.

Boson-sampling, proposed by Aaronson \& Arkhipov \cite{bib:AaronsonArkhipov10}, is a simple, but non-universal approach to implementing a specific quantum algorithm using quantum optics, which is strongly believed to be intractable on a classical computer. Whilst not universal for quantum computing, it has attracted much attention as it is significantly simpler than KLM, requiring only single-photon state preparation, passive linear optics, and photo-detection. Several elementary proof-of-principle experimental demonstrations have been performed \cite{bib:Broome20122012, bib:Crespi3, bib:Tillmann4, bib:Spring2}. For an elementary introduction to boson-sampling, see Gard \emph{et al.} \cite{bib:GardBSintro}.

Recently, Motes, Gilchrist, Dowling \& Rohde (MGDR) \cite{bib:MotesLoop} presented a scheme for efficient, universal boson-sampling using a fiber-loop architecture with time-bin encoding, which has frugal experimental requirements, and is highly scalable.  However, the scheme was only shown to be universal for boson-sampling, and does not allow universal quantum computation.

Here we show that the same architecture can be made universal for quantum computation with only a straightforward modification to the classical processing and control. The scheme requires only two embedded fiber-loops, three dynamically controlled beamsplitters, a single push-button photon source, and a single time-resolved photo-detector. These experimental requirements are fixed, irrespective of the size of the computation, limited only by fiber lengths (which scales polynomially with the number of qubits) and their respective loss rates. The scheme has only a single point of interference, making alignment of optical elements significantly more favorable than other proposals, such as bulk optics or waveguide implementations, which require simultaneously aligning $\mathrm{poly}(n)$ points of interference, $n$ being the number of optical modes.

The viability of time-bin encoding using a loop architecture has been demonstrated with recent quantum walk experiments by Schreiber \emph{et al.} \cite{bib:Schreiber10, bib:Schreiber12}. Humphreys \emph{et al.} \cite{bib:HumphreysSingleSpatialMode} presented a similar scheme for LOQC in a single spatial mode, based on using the polarization degree of freedom to `address' individual time-bins, and implement arbitrary LOQC transformations, which they showed was sufficient for universal LOQC. Additionally, they performed an elementary experimental demonstration of a controlled-phase (CZ) gate (a maximally entangling two-qubit gate, which is universal when combined with single-qubit operations) using their approach. Their scheme relied on dynamically controlled polarization rotations to perform time-bin addressing, and polarization-dependent temporal displacement operations (implemented via birefringent crystals) to interfere time-bins. The MGDR scheme has different experimental requirements, requiring only dynamically controlled beamsplitters to implement arbitrary linear optics interferometers.

%
%

\section{Boson-sampling}

\begin{figure}[!htb]
\includegraphics[width=0.9\columnwidth]{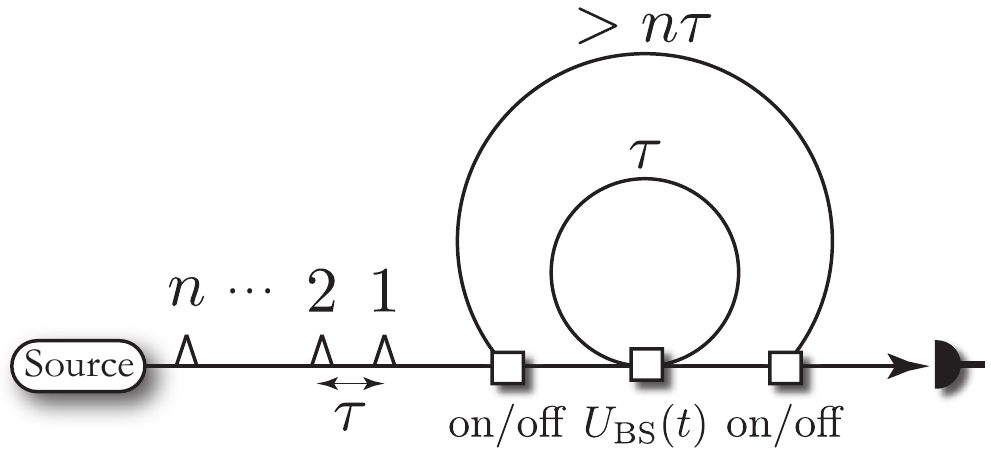}
\caption{The MGDR architecture for implementing arbitrary passive linear optics transformations on an $n$ time-bin-encoded pulse-train. The push-button source prepares a pulse train of single-photon and vacuum states, separated by time $\tau$, across $n$ time-bins. The pulse-train enters the outer loop, classically controlled via two on/off switches. The pulse-train then passes through the inner loop some number of times, as determined by the control of the central dynamically controlled beamsplitter. With \mbox{$\mathrm{poly}(n)$} round-trips of the outer loop, an arbitrary $n$-mode linear optics transformation may be implemented. The round-trip time of the inner loop is $\tau$, and the round-trip time of the outer loop is \mbox{$>n\tau$}. The first and last dynamically controlled beamsplitters need only be on/off (completely reflective or completely transmissive), whereas the central dynamically controlled beamsplitter must be able to implement arbitrary $SU(2)$ transformations.} \label{fig:full_arch}
\end{figure}

The full fiber-loop architecture for boson-sampling, as originally presented by MGDR, is shown in Fig. \ref{fig:full_arch}. A push-button single-photon source prepares a pulse-train of single-photon and vacuum states, where each time-bin is separated by time $\tau$. $\tau$ must be sufficiently large compared to the photons' wavepackets that they remain temporally orthogonal. The pulse-train enters a system of two embedded fiber-loops, controlled by three (classically) dynamically controlled beamsplitters. MGDR showed that this configuration, with appropriate control over the dynamically controlled beamsplitters, can implement an arbitrary, passive linear optics network, implementing a transformation on the mode operators of the form,
\begin{equation} \label{eq:creation_U}
\hat{a}_i^\dag \to \sum_{j=1}^n U_{i,j} \hat{a}_j^\dag,
\end{equation}
where $\hat{a}^\dag_i$ is the photon creation operator on the $i$th mode (time-bin), for any \mbox{$n\times n$} unitary $U$.

\begin{figure}[!htb]
\includegraphics[width=0.9\columnwidth]{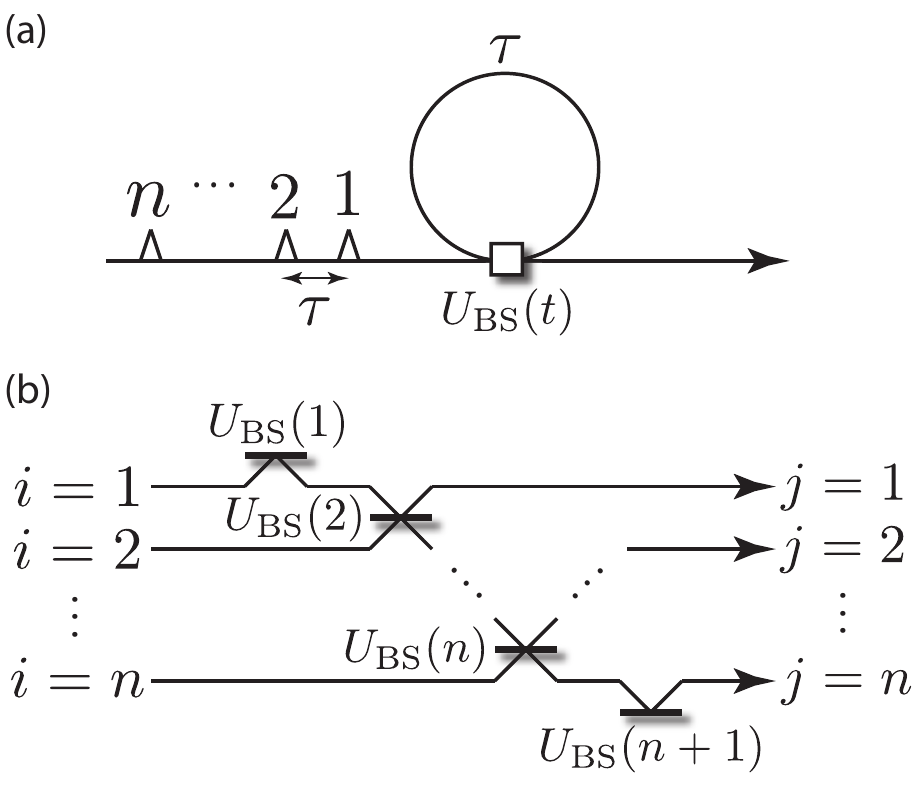}
\caption{Mapping between the inner fiber-loop (a) and its equivalent spatially-encoded transformation (b). $n$ passes through the inner loop is equivalent to a diagonal array of spatially-encoded beamsplitters. On its own, this configuration is not sufficient for arbitrary linear optics transformations as per Eq. \ref{eq:creation_U}.} \label{fig:single_loop}
\end{figure}

\begin{figure}[!htb]
\includegraphics[width=0.8\columnwidth]{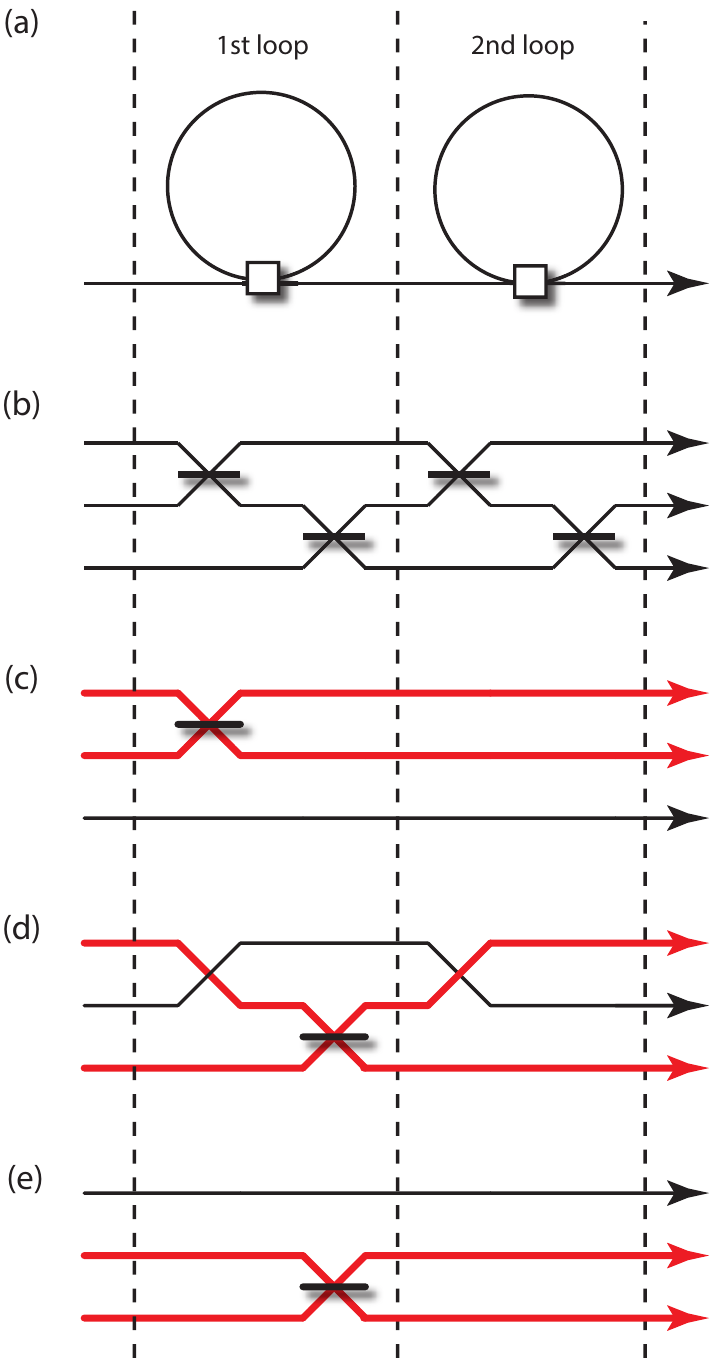}
\caption{(Color online) For \mbox{$n=3$} modes, with two consecutive applications of the inner loop (i.e two round-trips of the outer loop) and appropriate choices of beamsplitter reflectivities, arbitrary pairwise beamsplitter operations may be implemented: (a) between modes 1 and 2; (b) between modes 1 and 3; (c) between modes 2 and 3. It follows from the Reck \emph{et al.} decomposition that \mbox{$O(n^2)$} pairwise beamsplitter operations enables arbitrary 3-mode linear optics transformations.} \label{fig:two_loops}
\end{figure}

We begin by reviewing how this experimental configuration allows arbitrary transformations of the form of Eq. \ref{eq:creation_U} to be implemented. In Fig. \ref{fig:single_loop} we show the expansion for the inner loop as its equivalent spatially-encoded beamsplitter network with $n$ modes. In Fig. \ref{fig:two_loops} we show that with two passes through the inner loop, arbitrary pairwise beamsplitter transformations may be implemented in the case of \mbox{$n=3$} modes. These two passes through the inner loop are implemented via two round-trips through the outer loop.

\begin{figure}[!htb]
\includegraphics[width=0.7\columnwidth]{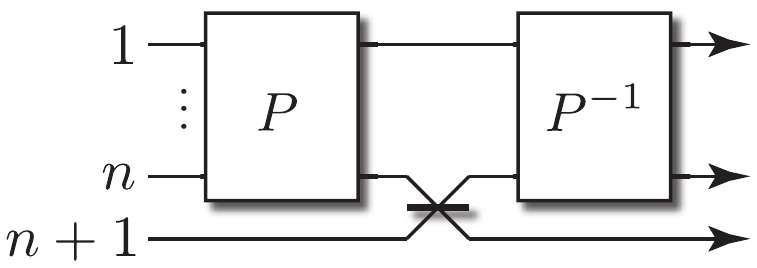}
\caption{Inductive argument that the MGDR architecture enables arbitrary $n$-mode linear optics transformations. We have established in Fig. \ref{fig:two_loops} that arbitrary 3-mode transformations are possible, which includes all permutations. To enable arbitrary $n$-mode transformations, we must additionally allow arbitrary beamsplitter operations between any of the first $n$ modes and the \mbox{$(n+1)$}th mode. We let $P$ be a permutation that maps any of the first $n$ modes to the $n$th mode, apply the desired beamsplitter interaction between the $n$th mode and the \mbox{$(n+1)$}th mode, and then apply the inverse permutation. Thus, if arbitrary $n$-mode transformations are possible, it follows inductively that arbitrary \mbox{($n+1$)}-mode transformations are possible, thus enabling universal linear optics transformations for arbitrary $n$.} \label{fig:induction}
\end{figure}

Having established that arbitrary pairwise beamsplitter operations are possible, it follows from the Reck \emph{et al.} \cite{bib:Reck94} decomposition that an arbitrary 3-mode linear optics network may be implemented with multiple applications of pairwise interactions. With arbitrary 3-mode linear optics networks at our disposal in the MGDR architecture, in Fig. \ref{fig:induction} we show via an inductive argument that with \mbox{$\mathrm{poly}(n)$} passes through the outer loop, arbitrary pairwise transformations between the $n$ modes are possible, and thus arbitrary $n$-mode linear optics transformations may be implemented using the Reck \emph{et al.} decomposition with $\mathrm{poly}(n)$ round-trips of the outer loop.

In the original proposal for boson-sampling presented by Aaronson \& Arkhipov, we require an input state comprising a \emph{specific} configuration of single-photon and vacuum states. With non-deterministic photon sources -- as is the case with most present-day technologies -- this might be achieved by multiplexing many heralded sources, as was considered by Motes \emph{et al.} \cite{bib:MotesSPDC}. However, recently Lund \emph{et al.} \cite{bib:RandomBS} showed that when using spontaneous parametric down-conversion (SPDC) as heralded sources, multiplexing is not necessary to implement a computational problem equivalent to boson-sampling. Rather, as long as the heralding signature is known, the exact heralding configuration does not affect the computational complexity. This is referred to as `randomized' boson-sampling. The MGDR architecture is perfectly suited to randomized boson-sampling, where we employ an SPDC with high repetition rate as the photon source preparing the pulse-train.

%
%

\section{Universal quantum computing}

We have established that the fiber-loop architecture may implement arbitrary passive linear optics transformations of the form of Eq. \ref{eq:creation_U}. Whilst this is sufficient for arbitrary boson-sampling on $n$-modes, which only requires passive linear optics, it is not sufficient for universal LOQC, which requires ancillary states, post-selection, and feed-forward. To demonstrate the universality of this scheme for LOQC, we will show that it can be made equivalent to various universal approaches for LOQC, with appropriate adjustments.

We will first show how the fiber-loop architecture can be mapped to the original KLM scheme for universal LOQC. Then we will show mappings to two subsequent variations employing cluster states -- by Nielsen \cite{bib:Nielsen04}, and Browne \& Rudolph \cite{bib:BrowneRudolph05} --  that substantially reduce experimental resource requirements.

In all of these variations, the necessary modification to the MGDR architecture is to introduce `ancilla injection' and `ancilla extraction', whereby we dynamically introduce ancillary photons into the photon pulse-train, or extract a subset of the photons in the pulse-train for measurement. Injecting and measuring these ancillary photons is the crucial ingredient necessary to make the scheme universal for LOQC, and does not require any experimental overhead compared to the original MGDR architecture, since the required switching to perform ancilla injection and extraction is already a part of the architecture. The only overhead compared to the original MGDR architecture is more complicated classical control of the dynamic switches, which requires feed-forward from the measurement results obtained during the ancilla extraction stage.

%
%

\subsection{Knill, Laflamme \& Milburn} \label{sec:KLM}

It was long believed that universal optical quantum computation would only be possible with strong Kerr non-linearities. In a seminal result, KLM showed that just linear optics transformations, photo-detection, and feed-forward, is sufficient for efficient universal quantum computation. The key observation was that, in conjunction with ancillary states (vacuum and single-photon Fock states), post-selection enables an effective non-linearity to be implemented. Specifically, a non-deterministic non-linear sign-shift (NS) gate may be implemented, which simulates a strong Kerr non-linearity. This gate subsequently allows for the construction of a non-deterministic CZ gate. Whilst non-deterministic, KLM presented an encoding scheme, based on gate teleportation \cite{bib:GottesmanChuang99}, which overcomes gate non-determinism, with only \mbox{$\mathrm{poly}(n)$} overhead, where $n$ is the number of logical qubits. Since the advent of the original KLM scheme, numerous proposals have been presented, which significantly improve on KLM's original scaling requirements, most promisingly using cluster state approaches \cite{bib:Raussendorf01, bib:Raussendorf03, bib:Nielsen04, bib:Nielsen06, bib:BrowneRudolph05, bib:Varnava05}.

The KLM scheme is efficient, requiring \mbox{$\mathrm{poly}(n)$} ancillary states and \mbox{$\mathrm{poly}(n)$} time-steps. Single-qubit operations are trivial using dual-rail encoding (whereby a qubit is encoded as a superposition of a single photon across two optical modes, \mbox{$\ket{\psi}=\alpha\ket{1,0}+\beta\ket{0,1}$}), requiring only arbitrary \mbox{$SU(2)$} beamsplitter operations. However, two-qubit entangling gates, such as the CZ gate, which are required to construct a universal gate set, require ancillary states, post-selection, and feed-forward.

\begin{figure}[!htb]
\includegraphics[width=0.5\columnwidth]{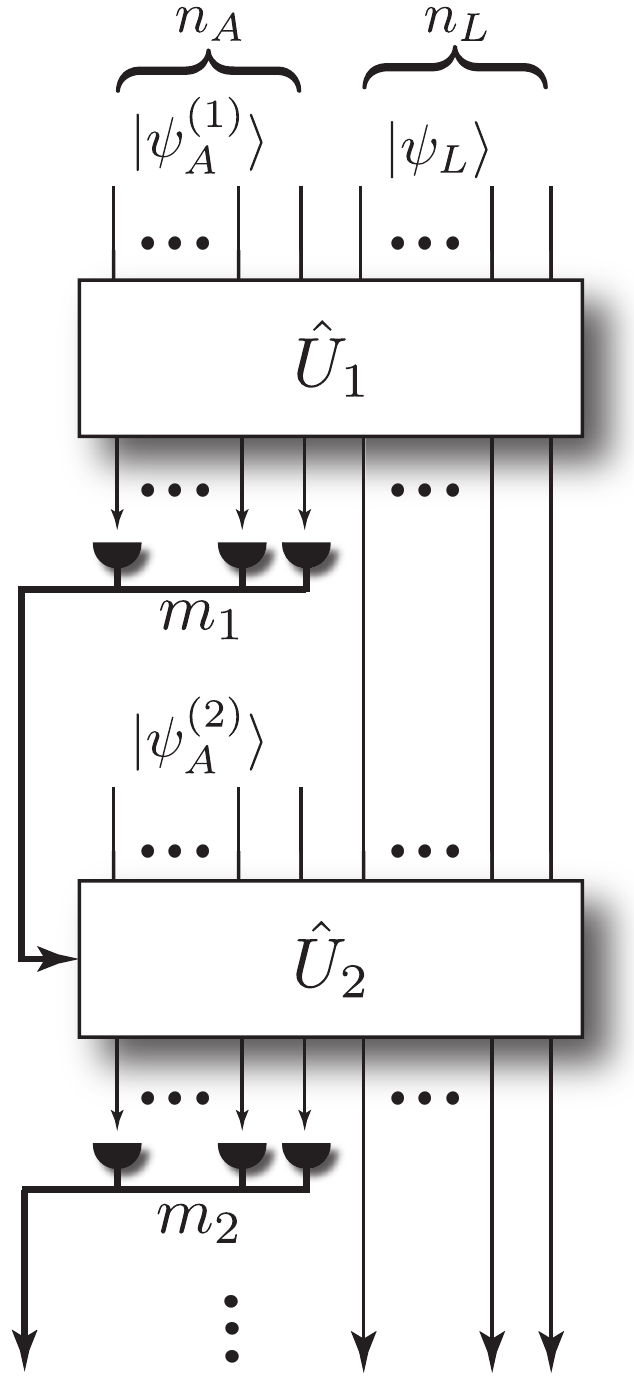}
\caption{A representation of the KLM scheme for universal LOQC. We prepare a logical state $\ket{\psi_L}$, comprising $n_L$ modes, and apply some \mbox{$\mathrm{poly}(n)$} number of iterations of: $n_A$-mode ancillary state preparation, $\ket{\psi_A^{(i)}}$, at the $i$th step; passive linear optics, $\hat{U}_i$; measurement of the ancillary state, yielding measurement signature $m_i$; and, feed-forward of $m_i$ to control the next passive unitary, $\hat{U}_{i+1}$. From KLM, this scheme enables efficient implementation of a universal gate set, enabling universal quantum computation.} \label{fig:KLM}
\end{figure}

The KLM scheme can be represented in the form shown in Fig. \ref{fig:KLM}, comprising logical state preparation, $\ket{\psi_L}$, and \mbox{$\mathrm{poly}(n)$} iterations of: ancillary state preparation, $\ket{\psi_A^{(i)}}$; photo-detection; and, passive linear optics, classically controlled by the previous measurement outcomes, $m_i$. This scheme is sufficient for universal LOQC, using \mbox{$\mathrm{poly}(n)$} circuit size. We will now show that the MGDR architecture can be made equivalent to the KLM scheme.

We have already determined that the MGDR architecture is sufficient for implementing all the $\hat{U}_i$ in Fig. \ref{fig:KLM}. The remaining requirement to demonstrate equivalence with KLM is that we are able to add arbitrary ancillary states prior to each $\hat{U}_i$, and measure them following $\hat{U}_i$, prior to the application of $\hat{U}_{i+1}$. We will consider these two stages separately, which we will refer to as `ancilla injection' and `ancilla extraction' respectively.

\begin{figure}[!htb]
\includegraphics[width=\columnwidth]{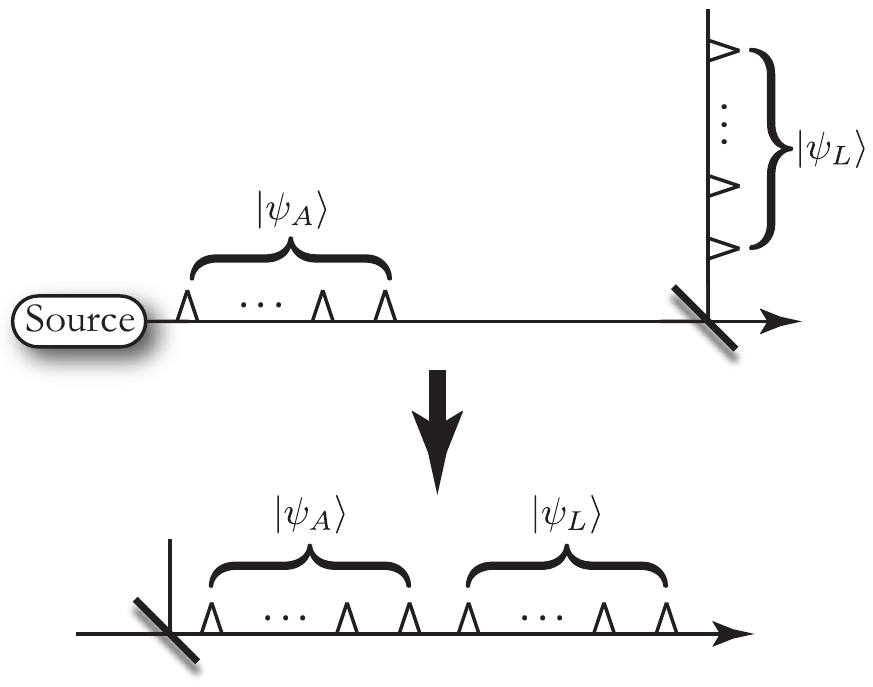}
\caption{The shown beamsplitter refers to the first beamsplitter in Fig. \ref{fig:full_arch}. Injection of ancillary state $\ket{\psi_A}$ into the outer loop is achieved by first setting the dynamic beamsplitter to be completely reflective for duration \mbox{$n_L\tau$}, which completely couples the logical state $\ket{\psi_L}$ into the outer loop. Then the beamsplitter is set to be completely transmissive, which completely couples the ancillary state $\ket{\psi_A}$ (prepared via the push-button source) into the outer loop. Immediately following the beamsplitter, we have a pulse-train of \mbox{$n_L+n_A$} time-bins, which combined represent the logical and ancillary states.} \label{fig:ancilla_injection}
\end{figure}

The protocol for ancilla injection is shown in Fig. \ref{fig:ancilla_injection}. This takes place at the first of the on/off dynamic switches in conjunction with the push-button single-photon source. The on/off switch is set to be completely reflective for duration \mbox{$n_L\tau$}, which completely couples all the logical time-bins into the the outer loop. Immediately following this, the switch is set to become completely transmissive for duration \mbox{$n_A\tau$}, thereby completely coupling the ancillary state $\ket{\psi_A^{(i)}}$, which is prepared via the push-button source, into the outer loop. Then the state immediately after the first switch is the logical state (across $n_L$ time-bins) followed by the ancillary state (across $n_A$ time-bins), yielding a combined pulse-train over \mbox{$n_L+n_A$} temporal modes. Following this, the MGDR protocol is applied to implement the required passive linear optics network $\hat{U}_i$.

\begin{figure}[!htb]
\includegraphics[width=0.85\columnwidth]{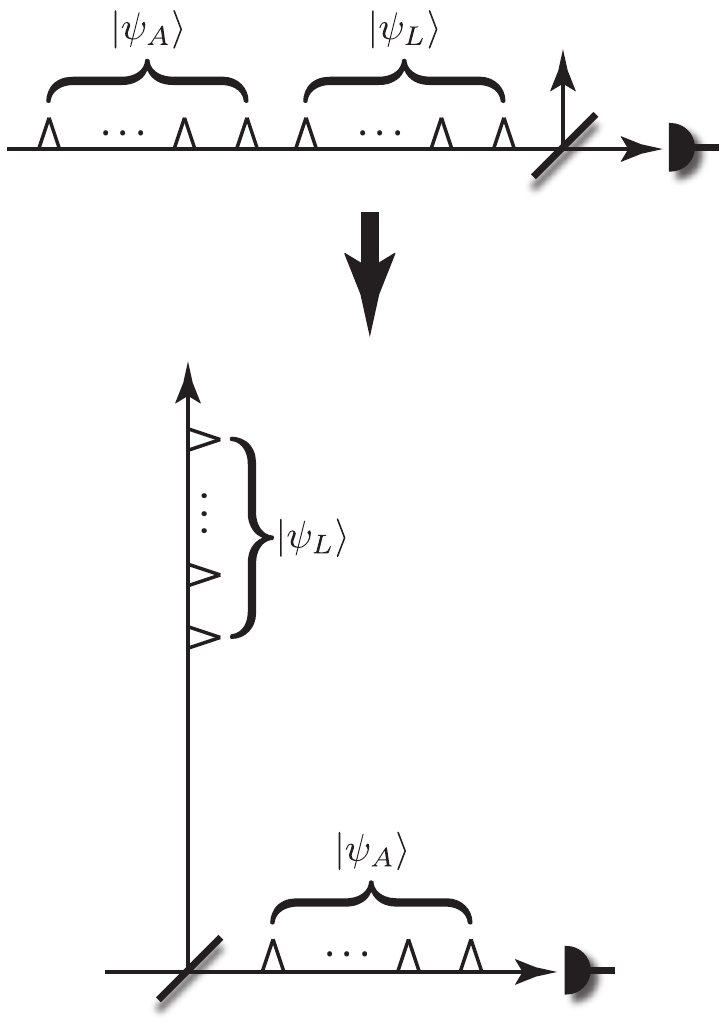}
\caption{The shown beamsplitter refers to the last beamsplitter in Fig. \ref{fig:full_arch}. For the first $n_L\tau$ time-steps, the beamsplitter is set to be completely reflective, which completely couples the logical state $\ket{\psi_L}$ into the outer loop. Following this, the beamsplitter is set to be completely transmissive, which completely couples the ancillary time-bins, $\ket{\psi_A}$, out of the outer loop and into the photo-detector. The measurement signature, $m_i$, is used to classically control the subsequent unitary $\hat{U}_{i+1}$.} \label{fig:ancilla_extraction}
\end{figure}

Next we must measure the ancillary modes (but not the logical modes) via ancilla extraction. This is the same as the ancilla injection protocol in reverse, as shown in Fig. \ref{fig:ancilla_extraction}. We set the last dynamic switch to be completely reflective for duration $n_L\tau$, completely coupling $\ket{\psi_L}$ into the outer loop. Following this, we set the switch to become completely transmissive, which couples the last $n_A$ time-bins, representing $\ket{\psi_A}$, out of the loop and into the photo-detector, yielding measurement signature $m_i$. Now only the logical modes remain inside the outer loop, until ancilla injection at the end of the round-trip. The measurement signature $m_i$ of the $n_A$ ancilla time-bins is used to classically control the passive linear optics transformation $\hat{U}_{i+1}$ at the subsequent step, which simply corresponds to a classical programming of the sequence of beamsplitter reflectivities in the central beamsplitter of Fig. \ref{fig:full_arch}.

Clearly, the protocol of: (1) initial logical state preparation, $\ket{\psi_L}$; (2) ancilla state preparation, $\ket{\psi_A^{(i)}}$, at every time-step $i$; (3) dynamically controlled passive linear optics, $\hat{U}_i$; (3) measurement of the ancillary modes, yielding measurement signatures $m_i$; and, (4) dynamic control of $\hat{U}_{i+1}$ at the subsequent step, is sufficient to enable the decomposition shown in Fig. \ref{fig:KLM}, enabling full KLM LOQC using \mbox{$\mathrm{poly}(n)$} physical resources. From KLM, using dual-rail encoding, the number of modes required to represent the logical state $\ket{\psi_L}$, scales as \mbox{$n_L=2n_Q$}, where $n_Q$ is the number of logical qubits. The number of ancillary states scales as \mbox{$n_A=\mathrm{poly}(n_L)$}. And thus the total number of photons and temporal modes in the system scales as $\mathrm{poly}(n_Q)$. Thus, the scheme is efficient.

%
%

\subsection{Cluster states}

`Cluster state' quantum computing \cite{bib:Raussendorf01, bib:Raussendorf03, bib:Nielsen06}, also known as `graph state quantum computing, `measurement-based quantum computing' and `one-way quantum computing', is an approach to universal quantum computing that differs from the usual and more familiar circuit-based approach. Rather than preparing an input state, applying some combination of single- and two-qubit gates belonging to a universal gate set, and measuring the output state, we instead prepare a `cluster state', which is a highly entangled multi-qubit state.

\begin{figure}[!htb]
\includegraphics[width=0.35\columnwidth]{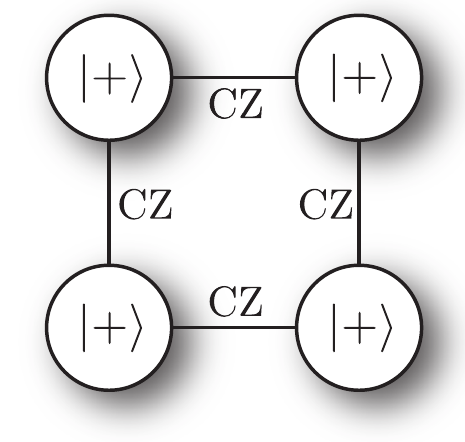}
\caption{A simple example of a four-qubit cluster state. Vertices represent qubits initialized in the $\ket{+}$ state, and edges represent the application of CZ gates. CZ gates commute, so the time ordering of the operations is irrelevant.} \label{fig:cluster_state}
\end{figure}

A cluster state may be represented as a graph, in which vertices represent qubits prepared in the \mbox{$\ket{+}=(\ket{0}+\ket{1})/\sqrt{2}$} state, and edges represent the application of CZ gates between the respective vertices. A simple example is shown in Fig. \ref{fig:cluster_state}. Having prepared this state, we have a `substrate' from which any quantum computation may be implemented using only single-qubit measurements, where the measurement bases are determined via classical feed-forward from previous measurement outcomes.

However, whilst a cluster state may be considered in terms of CZ gates acting on the edges of a graph, CZ gates are not the only way to prepare cluster states. For example, Bell pairs, which may be produced via a variety of means, such as SPDC, are locally equivalent to two-qubit cluster states.

The attractive feature of cluster states is that, having prepared the substrate state, an arbitrary quantum computation requires only single-qubit measurements. In a photonic context this is extremely attractive, as single-qubit measurements are trivial.

Cluster states have many other attractive properties. Gross \emph{et al.} \cite{bib:Gross06}, and Rohde \& Barrett \cite{bib:RohdeBarrett07} showed that cluster states may be prepared efficiently using non-deterministic gates, gate non-determinism being the bane of LOQC and the reason why KLM has such high experimental resource overhead. Additionally, cluster states were shown by Varnava \emph{et al.} \cite{bib:Varnava05} to be highly robust against qubit loss, and cluster states are directly suited to various topological codes, to achieve fault-tolerance with extremely favorable fault-tolerance thresholds.

Nielsen \cite{bib:Nielsen04} first demonstrated that by combining a non-deterministic KLM CZ gate with the cluster state formalism, physical resource requirements -- in terms of the number of required qubits and CZ gates -- could be reduced by orders of magnitude compared to KLM. Subsequently, Browne \& Rudolph \cite{bib:BrowneRudolph05} demonstrated that by replacing the KLM CZ gate with parity measurements, physical resource requirements could be further reduced substantially.

Next we will show that the fiber-loop architecture, combined with ancilla injection and extraction, can be mapped to both the Nielsen, and Browne \& Rudolph schemes, making the architecture suitable for highly efficient LOQC.

%
%

\subsubsection{Nielsen cluster states}

The Nielsen \cite{bib:Nielsen04} approach to optical cluster state quantum computing directly combines a non-deterministic KLM CZ gate with the cluster state formalism.

To illustrate the concept, let us consider the preparation of a two qubit cluster state (i.e a Bell pair). We prepare two $\ket{+}$ states, and apply the CZ gate, as described by KLM. KLM actually describe a hierarchy of CZ gates, with increasing success probability, but also with (polynomially) increasing resource overhead. Let us choose a particular such gate and let the gate success probability be $p_\mathrm{gate}$. Then, on average we will have to repeat the protocol \mbox{$1/p_\mathrm{gate}$} times to prepare a two-qubit cluster state.

\begin{figure}[!htb]
\includegraphics[width=0.75\columnwidth]{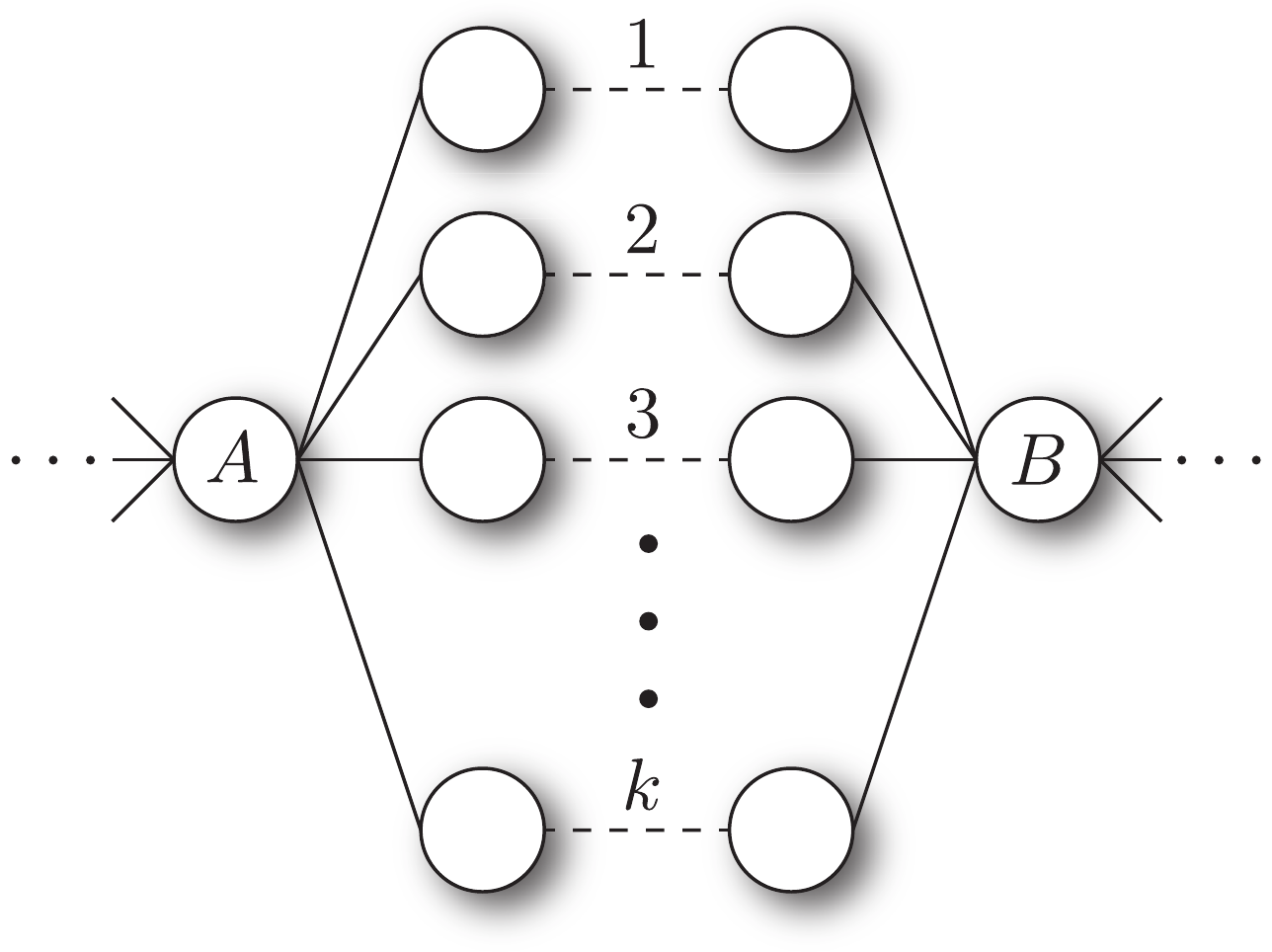}
\caption{An example of the Nielsen protocol for bonding two micro-clusters. We non-deterministically prepare two small clusters with central vertices $A$ and $B$, each with a series of $k$ branches emanating from the central vertex. To join the microclusters we perform non-deterministic CZ gates between the branches of the two micro-clusters in parallel. If a CZ gate fails it simply removes the respective branch qubits from the micro-clusters, and if it succeeds it creates an edge between the respective qubits. For a given gate success probability ($p_\mathrm{gate}$) and desired bonding success probability ($p_\mathrm{bond}$), the required number of branches $k$ is given by Eq. \ref{eq:nielsen_bond_suc}. Having successfully joined the two micro-clusters, all qubits other than $A$ and $B$ are measured in the $Y$ basis, which simply removes all those qubits, whilst creating edges between their neighboring qubits. Thus, after doing this to all qubits other than $A$ and $B$, we are left with just $A$ and $B$, with an edge between them. This idea generalizes to arbitrary graph topologies, whilst remaining efficient.} \label{fig:micro_cluster}
\end{figure}

The basic building block of the Nielsen protocol is to bond $\ket{+}$ states together to form star-shaped `micro-cluster' states, with $k$ vertices emanating from a central vertex. We perform this in advance, preparing a number of star clusters offline. The central vertex in each star cluster will ultimately belong to our final cluster state. The emanating branches facilitate joining multiple star clusters together. At the first instance, we take two stars and attempt to bond them together via their branches using the KLM CZ gate. When the CZ gate succeeds, we have prepared two `fused' stars. When it fails, the branches to which the CZ gate was applied are simply removed, leaving us with two star clusters with \mbox{$k-1$} branches, which may be reused for the next attempt at bonding the micro-clusters. However, with $k$ branches, we may have $k$ attempts to perform the joining of the stars. Thus, with
\begin{equation} \label{eq:nielsen_bond_suc}
k=\frac{\mathrm{log}(1-p_\mathrm{bond})}{\mathrm{log}(1-p_\mathrm{gate})}
\end{equation}
branches, we can successfully join two stars with success probability $p_\mathrm{bond}$. The bonding of two micro-clusters is shown in Fig. \ref{fig:micro_cluster}.

The protocol continues as one would expect, successively bonding larger and larger smaller clusters to form larger clusters, until we have a cluster of the size required for our computation. Using this protocol, arbitrarily large cluster states may be prepared efficiently, with resource requirements that Nielsen showed were substantially more favorable than the original KLM protocol.

In Sec. \ref{sec:KLM} we already established that the fiber-loop architecture is sufficient for universal KLM LOQC. Thus, the CZ gate required for Nielsen-style cluster state preparation is available to us. Additionally, we must manipulate the topology of the graph, depending on which CZ gates succeed or fail. Manipulating the topology of the graph is trivial, since it only requires permuting time-bins, and of course permutations on optical modes are unitary operations, and thus may be implemented using the MGDR protocol. Permutations only require completely transmissive or reflective beamsplitter operations in the MGDR architecture, and thus do not constitute major mode-matching problems.

Thus, by combining the MGDR protocol with ancilla injection and extraction, we are able to efficiently implement the cluster state approach of Nielsen, yielding a far more efficient approach to universal LOQC than simply using the fiber-loop architecture to implement KLM directly. 

%
%

\subsubsection{Browne \& Rudolph cluster states}

Whilst the Nielsen scheme is far more favorable than KLM, it nonetheless relies on the non-deterministic KLM CZ gate as its basic building block. Implementing this gate is challenging, as it is based on a Mach-Zehnder interferometer, which requires stabilization on the order of the photons' wavelength, and also requires a large number of optical elements. A far more favorable approach would be to employ only Hong-Ou-Mandel \cite{bib:HOM87}-type interference, which only requires stabilization on the order of the coherence length, which, from an experimental point of view, is extremely advantageous.

To this end, Browne \& Rudolph \cite{bib:BrowneRudolph05} proposed a scheme not requiring any Mach-Zehnder interference. Instead, their scheme relies on polarization encoded qubits (\mbox{$\ket{\psi}=\alpha_0\ket{H}+\alpha_1\ket{V}$}), where bonding smaller clusters into larger clusters is implemented via a joint parity measurement, which is easily implemented via a polarizing beamsplitter (PBS).

\begin{figure}[!htb]
\includegraphics[width=0.65\columnwidth]{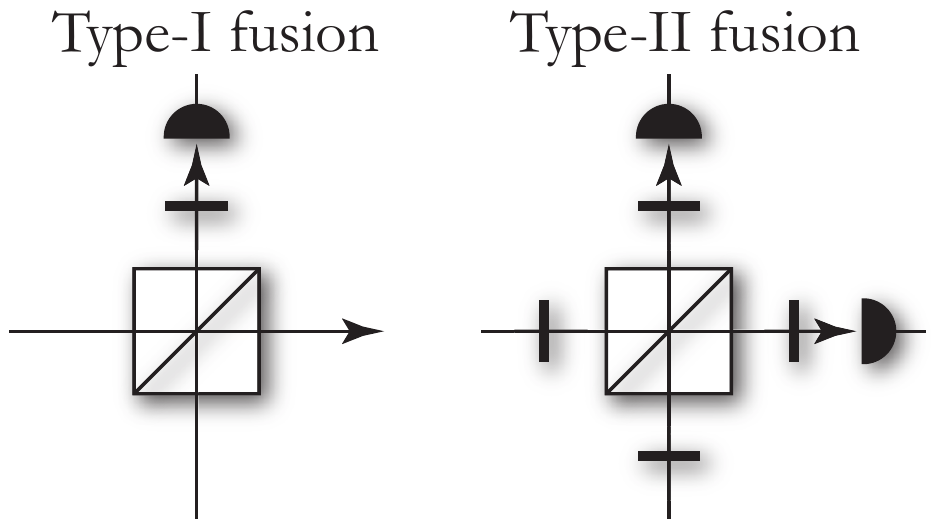}
\caption{The type-I and type-II fusion gates, comprising two polarization-encoded qubits, a PBS, waveplates, and polarization-resolved photo-detectors. The type-I fusion gate destroys a single qubit, whilst the type-II fusion gate destroys both qubits. Both gates require only Hong-Ou-Mandel-type interference, and thus are more favorable from a mode-matching perspective than the KLM CZ gate, which requires Mach-Zehnder interference.} \label{fig:fusion}
\end{figure}

Browne \& Rudolph's scheme employs three components: a resource of polarization-encoded Bell pairs; and, type-I and type-II `fusion' gates. The resource of Bell pairs may be prepared offline using a KLM gate (note that entangled pairs prepared via SPDC are \emph{not} suitable for their scheme, owing to the higher order photon-number terms). The type-I and type-II fusion gates are variations on parity measurement, implemented via a combination of waveplates and a PBS, shown in Fig. \ref{fig:fusion}. With these primitives at our disposal, we may prepare arbitrary cluster states with very high efficiency. Refer to Ref. \cite{bib:BrowneRudolph05} for full details of the scheme and its scaling characteristics. A detailed analysis of mode-matching effects in this scheme was presented by Rohde \& Ralph \cite{bib:RohdeRalph06}.

We will assume that a resource of Bell pairs is available (e.g prepared using KLM gates), and proceed to show that the remaining two requirements -- PBSs and waveplates -- may be mapped to the MGDR architecture, where polarization encoding is mapped to time-bin encoding.

In polarization encoding, a qubit takes the form \mbox{$\ket\psi = \alpha_0\ket{H} + \alpha_1\ket{V}$}, whereas in time-bin encoding the same state takes the form \mbox{$\ket{\psi} = \alpha_0\ket{1_t,0_{t+\tau}} + \alpha_1\ket{0_t,1_{t+\tau}}$}, where the qubit arrives at time $t$ and the time-bin separation is $\tau$.

First, let us consider the action of a waveplate in polarization encoding. A waveplate implements a polarization rotation such that $\vec\alpha' = \hat{U}\vec\alpha$, where \mbox{$\hat{U}\in SU(2)$}. Now we wish to find the equivalent operation in time-bin encoding. Time-bin encoding can be thought of as dual-rail encoding, where the two rails are temporally rather than spatially separated. In dual-rail encoding, the operation equivalent to a waveplate in polarization encoding is the beamsplitter operation. Thus, to implement a single-qubit rotation, we must implement an arbitrary beamsplitter between the temporal modes at $t$ and \mbox{$t+\tau$}. The MGDR protocol describes how to implement arbitrary pairwise beamsplitter operations between temporal modes.

\begin{figure}[!htb]
\includegraphics[width=0.65\columnwidth]{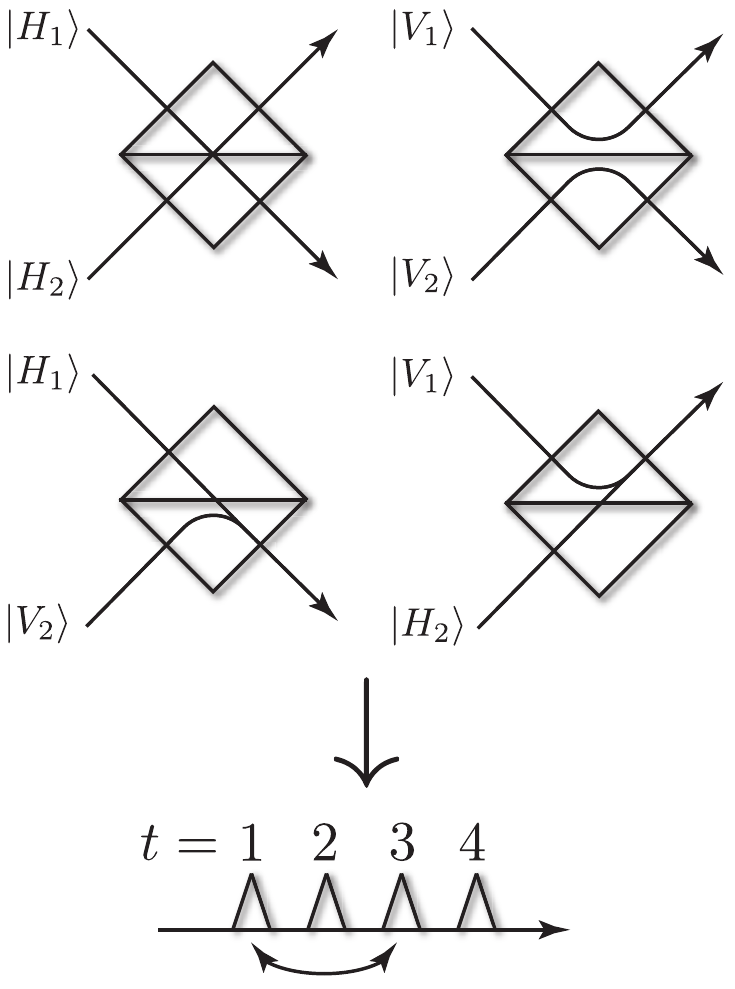}
\caption{(top) The action of a PBS on all four basis states of two polarization-encoded qubits. (bottom) The equivalent transformation in time-bin encoding based on the transformation of Eq. \ref{eq:PBS_mapping}, which is simply a swap of the 1st and 3rd time-bins.} \label{fig:PBS}
\end{figure}

Next, let us consider the PBS operation. In the two-mode polarization basis, the PBS operation implements the transformation,
\begin{eqnarray}
\hat{a}^\dag_{H_1} &\to& \hat{a}^\dag_{H_2}, \nonumber \\
\hat{a}^\dag_{V_1} &\to& \hat{a}^\dag_{V_1}, \nonumber \\
\hat{a}^\dag_{H_2} &\to& \hat{a}^\dag_{H_1}, \nonumber \\
\hat{a}^\dag_{V_2} &\to& \hat{a}^\dag_{V_2},
\end{eqnarray}
on the photonic creation operators. That is, horizontally polarized photons are transmitted, whilst vertically polarized photons are reflected. The full transformation on all four basis states is illustrated in Fig. \ref{fig:PBS}. This transformation may be represented in matrix form as,
\begin{equation} \label{eq:PBS_mapping}
\hat{U}_\mathrm{PBS} = \left(\begin{array}{cccc}
0 & 0 & 1 & 0 \\
0 & 1 & 0 & 0 \\
1 & 0 & 0 & 0 \\
0 & 0 & 0 & 1 \end{array}\right),
\end{equation}
in the basis $\{H_1,V_1,H_2,V_2\}$. Note that this is simply a permutation matrix that swaps the first and third time-bins, which of course is unitary, and thus may be implemented directly using the MGDR protocol. As discussed earlier, permutation operations do not require arbitrary beamsplitter operations -- we need only switch between completely transmissive and completely reflective at the central dynamically controlled beamsplitter -- thereby minimizing mode-matching effects.

To complete the type-I and type-II fusion gates we must measure some of the modes, which is accomplished via the ancilla extraction protocol.

Having demonstrated the ability of the fiber-loop architecture to implement type-I and type-II fusion gates, it follows from Browne \& Rudolph that arbitrary cluster states may be efficiently prepared, and their preparation strategies and scaling results apply directly.

As an interesting aside, using polarization encoding waveplates implement a polarization rotation, which does not generate entanglement, whereas the PBS is the entangling operation. On the other hand, in time-bin encoding the equivalent of the waveplate is the time-bin beamsplitter operation, which generates temporal entanglement, whereas the equivalent of the PBS -- a swap operation on time-bins -- does not. Nonetheless, despite the reversed role of entanglement generation, the two encodings are isomorphic.

%
%

\section{Conclusion}

We have presented a simple experimental architecture for universal LOQC using constant experimental resources. The scheme requires a single push-button photon source or Bell pair source, three dynamically controlled beamsplitters, two nested fiber-loops, and a single time-resolved photo-detector. The experimental requirements for the architecture are fixed, irrespective of the size of the computation, yet can efficiently implement full, universal LOQC, equivalent to the KLM protocol or subsequent and more efficient cluster state protocols. There is a single point of interference in the architecture, which may make optical alignment in this scheme far more favorable than existing bulk optics or waveguide approaches. The size of the computation that can be implemented is limited only by the length of the outer fiber-loop, which must have round-trip time of at least \mbox{$n\tau$}, where $n$ is the number of time-bins, which scales polynomially with the number of logical qubits in the computation. By removing the ancilla injection and ancilla extraction stages of the protocol, this scheme reduces to the MGDR architecture for universal boson-sampling. The required technologies to implement this scheme are, for the larger part, available today. Thus, elementary demonstrations of this protocol might be viable in the near future. 

\begin{acknowledgments}
We thank Keith Motes, Alexei Gilchrist, Timothy Ralph, Geoff Pryde, Andrew White, and Jonathan Dowling for helpful discussions. We thank Yoram Zekri for helping to motivate this work. This research was conducted by the Australian Research Council Centre of Excellence for Engineered Quantum Systems (Project number CE110001013).
\end{acknowledgments}

\bibliography{bibliography}

\end{document}